\documentclass[]{revtex4}

\usepackage{amssymb, amsmath, amsfonts, latexsym, verbatim, mathrsfs}
\usepackage{amssymb, amsmath, amsfonts, latexsym, verbatim, mathrsfs}
\usepackage{graphicx}

\begin{document}

\title{Indirect control of the asymptotic states of a quantum dynamical semigroup}

\author{Raffaele Romano}
\thanks{The author acknowledges support from the European grant
ERG:044941-STOCH-EQ. Work in part supported by Istituto Nazionale di
Fisica Nucleare, Sezione di Trieste, Italy.}

\affiliation{The Abdus Salam International Centre for Theoretical
Physics \\ Strada Costiera, 11 I-34014 Trieste, Italy}

\affiliation{The Department of Theoretical Physics, University of
Trieste, \\
Strada Costiera, 11 I-34014 Trieste, Italy}

\hyphenation{e-qui-va-lent}

\newtheorem{theorem}{Theorem}



\begin{abstract}

\noindent In the dynamics of open quantum systems, the interaction
with the external environment usually leads to a contraction of the
set of reachable states for the system as time increases, eventually
shrinking to a single stationary point. In this contribution we
describe to what extent it is possible to modify this asymptotic
state by means of indirect control, that is by using an auxiliary
system coupled to the target system in order to affect its dynamics,
when there is a purely dissipative coupling between the two systems.
We prove that, also in this restrictive case, it is possible to
modify the asymptotic state of the relevant system, give necessary
conditions for that and provide physical examples. Therefore, in
indirect control schemes, the environmental action has not only a
negative impact on the dynamics of a system, it is rather possible
to make use of it for control purposes.

\end{abstract}

\maketitle


\section*{Introduction}

The study of quantum mechanical systems is relevant not only for a
deep understanding of the fundamental physical laws, but also for
its potential applications \cite{niel}. In particular, it has been
proved that the use of quantum-based technologies would highly
increase the performance of a computational device \cite{shor}. This
is due to the mathematical structure of the theory, leading to
peculiar features for the microscopic systems, absent in the
macroscopic, classical world. Among them, the most relevant (and not
completely understood) is represented by the quantum correlations
known as {\it entanglement}, typical of quantum systems, whose
complete characterization has been given only in the low dimensional
cases. These correlations are the key of many recently proposed
protocols, as teleportation \cite{benn1} or quantum cryptography
\cite{benn2}.

However, quantum systems are fragile: their relevant features are
usually degraded by their interaction with the external environment,
leading to irreversibility, dissipation and decoherence. Therefore
closed systems, described by the Schr\"{o}dinger equation, are only
approximations of real systems, that are necessarily open since they
exchange energy and information with the external world. To account
for this, the standard dynamical model is given by a {\it quantum
dynamical semigroup}, that is a one-parameter family of Markovian
(i.e., satisfying the semigroup property) completely positive maps,
transforming pure states into mixtures and destroying quantum
coherence \cite{alic,breu}. Irreversibility is due to the fact that
quantum dynamical semigroups are contractions in the state space of
the system: this highly reduces the ability of manipulation on the
system, since many transitions become forbidden. The extremal case
is represented by the so-called {\it uniquely relaxing semigroups},
where there is a unique asymptotic state for every initial state of
the system. Therefore, in order to fully make use of the
potentialities of quantum systems, it is fundamental to understand
the mechanisms leading to dissipation and to conceive methods to
counteract them, and, more in general, to control the dynamics of
the system.

With these motivations in mind, several solutions to the problem
have been proposed. The study of the symmetries of the
system-environment coupling has lead to the notion of {\it
decoherence-free subspaces or subsystems}, that are quiet places,
unaffected by decoherence, where to encode the relevant information
(for a review on the topic, see \cite{lida}). From a more active
perspective, a theory of quantum control has developed, dealing with
the effect of external manipulations that can be performed on the
system. In the standard setting ({\it coherent control}), the
control parameters enter the Hamiltonian of the system, for example
through external fields coupled to the system (for a geometric
control perspective, in line with that developed in this work, refer
to \cite{schi,albe,viol1,viol2,tarn,rama,alta}). Although this is
the most natural way to introduce external actions in the dynamics
of a system, its ability to fight the unwished effects of the
irreversible dynamics is limited \cite{alta}, unless some
information about the state of the system is collected, and then
used to update in real time the controls. This information is
usually obtained via an indirect continuous measurement and, because
of this, the master equation describing the system becomes
stochastic. This {\it quantum feedback} scheme represents a
promising approach in many interesting cases \cite{bela,wise,manc}.

A different approach to quantum control has been recently discussed,
in which the control does not enter through the Hamiltonian of the
system, but it is rather obtained by means of an auxiliary system
(ancilla) \cite{mand,roma1,fu}, that can be manipulated and put in
interaction with the relevant system, and finally discarded at the
end of the procedure. The ability of driving the target system
through the auxiliary one is determined by the correlations between
them. The interest in this {\it indirect control} method is twofold.
It is complementary to the coherent control approach, that is, it
can be applied to experimental setups where the coherent control
technique is not appropriate. Moreover, in the indirect control
approach the environment does not only represent a source of noise,
it can also be used for control purposes. In fact, it has been
proved that the environment can correlate two systems immersed in it
via a noisy mechanism, not only destroy their existing correlations
\cite{brau,bena1}. This mechanism can be used to obtain total
control of the target system even if there is not a direct
interaction between the two parties \cite{roma2}.

In this contribution we prove that, in the indirect control scheme,
the environment induced correlations can be used to manipulate the
asymptotic states of the target system. This result gives further
evidence that, in this framework, it is possible to engineer the
environmental action to get controllability. The work is organized
as follows: in Section \ref{sec1} we summarize some standard results
about quantum dynamical semigroups and their stationary states, in
Section \ref{sec2} we provide necessary conditions for the indirect
manipulation of the asymptotic states of the relevant system under a
purely dissipative dynamics, in Section \ref{sec3} we evaluate the
stationary states and describe a concrete physical example of
application, and finally we conclude in Section \ref{sec4}.


\section{Quantum dynamical semigroups and their stationary states}
\label{sec1}

In many situations (usually, when there is a weak interaction with
the surrounding environment) it is possible to approximate the
reduced dynamics of an open system $S$ using a Markovian
one-parameter family of completely positive maps $\{ \gamma_t; t
\geqslant 0\}$, satisfying the semigroup property $\gamma_{t + s} =
\gamma_t \circ \gamma_s$, with $t, s \geqslant 0$, with
\begin{equation}\label{eq00}
    \rho_s (t) = \gamma_t [\rho_s (0)],
\end{equation}
where the state of the system $S$ is given by the Hermitian,
positive, unit trace operator $\rho_s$ (statistical operator),
acting on the $n$-dimensional Hilbert space associated to $S$.
Complete positivity is necessary in order to guarantee a physically
consistent interpretation of the formalism when dealing with
composite, entangled systems. The generator $L$ of the dynamics is
defined by $\dot{\rho}_s = L[\rho_s]$, and it is possible to prove
that it has the general form (the so-called Lindblad-Kossakowski
form)
\begin{equation}\label{eq01}
    L[\rho_s] = -i [H_s, \rho_s] + \sum_{i,j} c_{ij} \Bigl(
    F_i \rho_s F_j^{\dagger} - \frac{1}{2} \{ F_j^{\dagger} F_i, \rho_s \}
    \Bigr),
\end{equation}
where $H_s = H_s^{\dagger}$ is the Hamiltonian of $S$, and the set
$\{F_i; i = 1, \ldots, n^2 - 1 \}$, along with the $n$-dimensional
identity operator, form a basis for the operators acting on the
Hilbert space associated to $S$, satisfying ${\rm Tr} F_i = 0$, and
${\rm Tr} (F_i F_j^{\dagger}) = \delta_{ij}$. The $(n^2 - 1) \times
(n^2 - 1)$ matrix $C = [c_{ij}]$ (Kossakowski matrix) fulfills
$C^{\dagger} = C$ and $C \geqslant 0$, necessary and sufficient
condition for the complete positivity of the dynamics
\cite{alic,breu}.

In the following, $S$ is a bipartite system, $S = T + A$, where $T$
is the target system (to be manipulated) and $A$ the ancilla. We
assume that $T$ and $A$ are two copies of the same two-level system,
separately interacting with the same environment, assumed to be
spatially homogeneous, according to the Markovian dynamics
(\ref{eq01}). We further assume that $H_s = 0$, since we want to
study a purely dissipative dynamics. To model this system, it is
sufficient to consider the basis $\{F_i; i = 1, \ldots, 6 \}$, given
by the local operators $F_i = \sigma_i \otimes \mathbb{I}$ for $i =
1, 2, 3$ and $F_i = \mathbb{I} \otimes \sigma_{i-3}$ for $i = 4, 5,
6$, where $\sigma_i$, $i = 1, 2, 3$ are the Pauli matrices. We
consider the standard representation of these operators in which
$\sigma_3$ is diagonal. The $6 \times 6$ matrix $C$ has the form
\begin{equation}\label{eq03}
    C = \left[%
\begin{array}{cc}
  A & B \\
  B^{\dagger} & A \\
\end{array}%
\right],
\end{equation}
where $A = A^{\dagger}$ is the Kossakowski matrix for the system $T$
(or $A$) alone, and $B$ represents the dissipative coupling between
the two parties. $A$ and $B$ are $3 \times 3$ blocks. The form
(\ref{eq03}) is not the most general joint Kossakowski matrix, as we
have assumed that the two parties interact separately with the
environment, and that the two local dissipative contributions are
equal (homogenous environment). We will limit our attention to
models well described by (\ref{eq03}); moreover, for simplicity, we
will further assume $B = B^{\dagger}$. This choice produces a
significative simplification in the treatment and it is still of
great phenomenological interest.

The first term in the right hand side of (\ref{eq01}) represents the
coherent part of the evolution and it generates a group of
reversible, unitary transformations. The second term generates the
irreversible dynamics, according to the matrix $C$ whose entries
depend on the microscopical details of the interaction between
system and environment. It also leads to the appearance of
attractors in the state space of $S$, and consequently relaxation to
equilibrium of the states of the open system. A stationary state for
the dynamics, $\rho^{\infty}_s$, is defined by the condition on the
generator $L[\rho_s^{\infty}] = 0$. Since the dynamics is linear in
the state $\rho_s$, it is possible to fully characterize the
asymptotic fate of the system by studying the eigenvalues of the
dynamical matrix appearing in the coherence vector representation of
(\ref{eq01}) \cite{lend}. Although this treatment is very general,
it is not suitable for the purposes of this work. We will rather
refer to some necessary conditions for the existence of stationary
states, and for the convergence of $\rho_s (t)$ to them, given in
terms of the operators $\{ V_i; i \}$ appearing in the diagonal form
of (\ref{eq01}),
\begin{equation}\label{eq02}
    L[\rho_s] = \sum_{i} \Bigl(
    V_i \rho_s V_i^{\dagger} - \frac{1}{2} \{ V_i^{\dagger} V_i, \rho_s \}
    \Bigr).
\end{equation}
These conditions are expressed by the following theorem \cite{frig},
that has been adapted to the present context.

\begin{theorem}
\label{theo1} Given the quantum dynamical semigroup (\ref{eq02}),
assume that it admits a stationary state $\rho_0$ of maximal rank.
Defining $\mathcal{M} = \{ V_i, V_i^{\dagger}; i\}^{\prime}$, the
commutant of the Hamiltonian plus the dissipative generators, the
following conditions hold true:

    1. If $\mathcal{M} = {\rm span} (\mathbb{I})$, then $\rho_0$ is the
    unique stationary state. Moreover, if $\{ V_i; i \}$ is a self-adjoint
    set with $\{ V_i; i \}^{'} = {\rm span} (\mathbb{I})$, then for every
    initial condition $\rho_s (0)$
    $$\lim_{t \rightarrow + \infty} \rho_s (t) = \rho_0.$$

    2. If $\mathcal{M} \ne {\rm span} (\mathbb{I})$, then there exist a
    complete family $\{ P_n; n \}$ of pairwise orthogonal projectors
    such that $\mathcal{Z} = \mathcal{M} \cap \mathcal{M}^{'} = \{ P_n; n
    \}^{''}$. If $\{ V_i; i \}^{'} = \mathcal{M}$, two extreme cases together
    with their linear superpositions may occur.

    i. If $\mathcal{Z} = \mathcal{M}$, then for every initial condition $\rho_s (0)$
    $$\lim_{t \rightarrow + \infty} \rho_s (t) = \sum_n {\rm Tr} (P_n \rho_s (0) P_n)
    \frac{P_n \rho_0 P_n}{{\rm Tr} (P_n \rho_0 P_n)}.$$

    ii. If $\mathcal{Z} = \mathcal{M}^{'}$, then for every $\rho_s (0)$
    $$\lim_{t \rightarrow + \infty} \rho_s (t) = \sum_n P_n \rho_s (0) P_n.$$
\end{theorem}

Therefore, the stationary states of a quantum dynamical semigroup
can be characterized by means of the algebras $\mathcal{M}$,
$\mathcal{M}^{\prime}$, and $\mathcal{Z}$, if a maximal rank
stationary state $\rho_0$ is available. These quantities are
evaluated in the next section, depending on the form of the matrix
$C$.


\section{Relevant algebraic quantities}
\label{sec2}

Following Theorem \ref{theo1}, we need to write $C$ in diagonal form
in order to find the operators $V_i$ appearing in (\ref{eq02}). This
is achieved by means of the unitary transformation $U$ such that
\begin{equation}\label{eq04}
    U C U^{\dagger} = {\rm diag} (\lambda_i, i = 1, \ldots, 6),
\end{equation}
where $\lambda_i$ are the eigenvalues of $C$. $U$ has the form
\begin{equation}\label{eq05}
    U = \frac{1}{\sqrt{2}} \left[%
\begin{array}{cc}
  \tilde{U} & \tilde{U} \\
  - \hat{U} & \hat{U} \\
\end{array}%
\right]
\end{equation}
and $\tilde{U}$, $\hat{U}$ are unitary transformations such that
\begin{eqnarray}\label{eq06}
    \tilde{U} (A + B) \tilde{U}^{\dagger} &=& {\rm diag} (\lambda_i^+, i = 1, 2,
    3), \nonumber \\
    \hat{U} (A - B) \hat{U}^{\dagger} &=& {\rm diag} (\lambda_i^-, i = 1, 2,
    3).
\end{eqnarray}
The eigenvalues of $C$ are ordered as $\lambda_i = \lambda^+_i$ for
$i = 1, 2, 3$ and $\lambda_i = \lambda^-_{i-3}$ for $i = 4, 5, 6$.
Comparing the generator forms (\ref{eq01}) and (\ref{eq02}), and
using the notation $U = [u_{ij}]$, we have
\begin{equation}\label{eq07bis}
V_i = \sqrt{\lambda_i} \sum_{k = 1}^6 u_{ik}^* F_k.
\end{equation}
Following (\ref{eq05}), it is possible to write
\begin{equation}\label{eq07}
    \frac{1}{\sqrt{\lambda_i}} V_i = \left\{%
\begin{array}{ll}
    \mathbb{I} \otimes \tilde{\sigma}_i + \tilde{\sigma}_i \otimes \mathbb{I}, &
    \hbox{{\it i} = 1, 2, 3} \\ \\
    \mathbb{I} \otimes \hat{\sigma}_{i - 3} - \hat{\sigma}_{i - 3} \otimes \mathbb{I},
    \quad & \hbox{{\it i} = 4, 5, 6} \\
\end{array}%
\right.
\end{equation}
where we have defined
\begin{equation}\label{eq08}
    \tilde{\sigma}_i = \sum_{k = 1}^3 \tilde{u}^*_{ik} \sigma_k, \quad \quad
    \hat{\sigma}_i = \sum_{k = 1}^3 \hat{u}^*_{ik} \sigma_k,
\end{equation}
and we used the notation $\tilde{U} = [\tilde{u}_{ij}]$, $\hat{U} =
[\hat{u}_{ij}]$. The operators in (\ref{eq08}) satisfy ${\rm Tr} \,
\tilde{\sigma}_i = {\rm Tr} \, \hat{\sigma}_i = 0$ and ${\rm Tr}
(\tilde{\sigma}_i \tilde{\sigma}_j^{\dagger}) = {\rm Tr}
(\hat{\sigma}_i \hat{\sigma}_j^{\dagger}) = \delta_{ij}$. They are
self-adjoint if and only if the unitary operators $\tilde{U}$ and
$\hat{U}$ are orthogonal.

\noindent The commutant of Theorem \ref{theo1} can be expressed as
\begin{equation}\label{eq09}
    \mathcal{M} = \{ V_i, V_i^{\dagger}; i \vert \lambda_i \ne 0 \}^{'} =
    \bigcap_{i \vert \lambda_i \ne 0} \{ V_i, V_i^{\dagger} \}^{'},
\end{equation}
where only non-vanishing eigenvalues $\lambda_i$ have to be
considered, otherwise the corresponding $V_i$ do not appear in the
generator (\ref{eq02}). Moreover, for a given $i$,
\begin{equation}\label{eq10}
    \{ V_i, V_i^{\dagger} \}^{'} = \{ v \vert v \in \{ V_i \}^{'}, v^{\dagger}
    \in \{ V_i \}^{'} \},
\end{equation}
therefore we can limit our attention to the sets $\{ V_i \}^{'}$. We
find convenient to consider separately the two kinds of
contributions defined in (\ref{eq07}). To begin with, we consider a
fixed index $i$ such that $\lambda_i^+ \ne 0$, and assume that the
corresponding $\tilde{\sigma}_i$ is non-singular. In this case it
can be written as
\begin{equation}\label{eq12}
    \tilde{\sigma}_i = \tilde{\mu}_i R_i
    \sigma_3 R_i^{-1}
\end{equation}
where it is possible to choose $R_i = R_i^{-1}$, and
\begin{equation}\label{eq13bis}
    \tilde{\mu}_i^2 = \sum_j (\tilde{u}^*_{ij})^2.
\end{equation}
Since $\mathbb{I} \otimes \tilde{\sigma}_i + \tilde{\sigma}_i
\otimes \mathbb{I} = \tilde{\mu}_i \mathcal{R}_i ( \mathbb{I}
\otimes \sigma_3 + \sigma_3 \otimes \mathbb{I} ) \mathcal{R}_i$,
with $\mathcal{R}_i = R_i \otimes R_i$, it follows that
\begin{equation}\label{eq14}
    \{\mathbb{I} \otimes \tilde{\sigma}_i + \tilde{\sigma}_i \otimes
    \mathbb{I}\}^{'} = \mathcal{R}_i \{\mathbb{I} \otimes \sigma_3 + \sigma_3 \otimes
    \mathbb{I}\}^{'} \mathcal{R}_i
\end{equation}
and then, after the explicit computation,
\begin{equation}\label{eq15}
    \{ V_i \}^{'} = {\rm span} (\mathbb{I} \otimes \mathbb{I},
    \mathbb{I} \otimes \tilde{\sigma}_i, \tilde{\sigma}_i \otimes
    \mathbb{I}, \tilde{\sigma}_i \otimes \tilde{\sigma}_i, \Omega^+, \Delta_i^-),
\end{equation}
having defined the additional operators
\begin{eqnarray}\label{eq16}
    \Omega^+ &=& \sigma_1 \otimes \sigma_1 + \sigma_2 \otimes \sigma_2 + \sigma_3 \otimes
    \sigma_3, \nonumber \\
    \Delta_i^- &=& \mathcal{R}_i (\sigma_1 \otimes \sigma_2 -  \sigma_2 \otimes
    \sigma_1) \mathcal{R}_i.
\end{eqnarray}
Notice that, in general, the operators in the right hand side of
(\ref{eq15}) are not self-adjoint, nor orthogonal each other in the
Hilbert-Schmidt metric, since the transformation $\mathcal{R}_i$ is
not unitary (equivalently, self-adjoint). However, if the
coefficients $\tilde{u}^*_{ij}$, $j = 1, 2, 3$, are real,
$\tilde{\sigma}_i$ is self-adjoint and $\mathcal{R}_i$ unitary.
Consequently, in this case the basis of $\{ V_i \}$ is made of
Hermitian, orthogonal operators.

The commutants $\{ V_i \}^{'}$ are completely characterized for $i =
1, 2, 3$. Finally, $\{ V_i, V_i^{\dagger} \}^{'}$ can be found by
considering (\ref{eq10}):
\begin{equation}\label{eq15bis}
    \{ V_i, V_i^{\dagger} \}^{'} = \left\{%
\begin{array}{ll}
    \{ V_i \}^{'}, & \hbox{iff $\tilde{\sigma}_i = \tilde{\sigma}_i^{\dagger}$;} \\ \\
    {\rm span} (\mathbb{I} \otimes \mathbb{I}, \Omega^+), \quad & \hbox{otherwise.} \\
\end{array}%
\right.
\end{equation}

The corresponding sets for $i = 4, 5, 6$ can be found by applying
the same procedure to $\hat{\sigma}_i$, assuming that $\lambda^-_i
\ne 0$. The result is
\begin{equation}\label{eq18}
    \{ V_i \}^{'} = {\rm span} (\mathbb{I} \otimes \mathbb{I},
    \mathbb{I} \otimes \hat{\sigma}_i, \hat{\sigma}_i \otimes
    \mathbb{I}, \hat{\sigma}_i \otimes \hat{\sigma}_i, \Omega^-_i, \Delta_i^+),
\end{equation}
where
\begin{eqnarray}\label{eq18bis}
    \Omega^-_i &=& \mathcal{S}_i (\sigma_1 \otimes \sigma_1 - \sigma_2 \otimes
    \sigma_2) \mathcal{S}_i, \nonumber \\
    \Delta_i^+ &=& \mathcal{S}_i (\sigma_1 \otimes \sigma_2 + \sigma_2 \otimes
    \sigma_1) \mathcal{S}_i,
\end{eqnarray}
and $\mathcal{S}_i = S_i \otimes S_i$, with
\begin{equation}\label{eq18bbis}
    \hat{\sigma}_i = \hat{\mu}_i S_i
    \sigma_3 S_i^{-1},
\end{equation}
where $S_i = S_i^{-1}$, and
\begin{equation}\label{eq13tris}
    \hat{\mu}_i^2 = \sum_j (\hat{u}^*_{ij})^2.
\end{equation}
Finally, in this case
\begin{equation}\label{eq18tris}
    \{ V_i, V_i^{\dagger} \}^{'} = \left\{%
\begin{array}{ll}
    \{ V_i \}^{'}, & \hbox{iff $\hat{\sigma}_i = \hat{\sigma}_i^{\dagger}$;} \\ \\
    {\rm span} (\mathbb{I} \otimes \mathbb{I}), \quad & \hbox{otherwise.} \\
\end{array}%
\right.
\end{equation}

If $\tilde{\sigma}_i$ (or $\hat{\sigma}_i$) is singular, the
previous computations are not longer valid. In this case, the
commutants must be evaluated by direct computation and it is not
possible, in general, to express their structure in a compact form.

We have all the ingredients to evaluate the contribution related to
the dissipative generators $V_i$ in (\ref{eq09}). We find convenient
to denote by $n_+$ and $n_-$ the number of non-vanishing eigenvalues
of the type $\lambda^+$ and $\lambda^-$ respectively. The
non-trivial cases are summarized below, with the corresponding
relevant algebras and set of projectors $\{ P_i; i \}$, described in
theorem \ref{theo1}, to be used to construct the set of stationary
states. For further reference, the projectors $\Pi_k$, $k \in \{-,
+, 1, \ldots, 4 \}$, are defined as
\begin{eqnarray}\label{eq23bis}
    \Pi_k = [\pi^k_{ij}], \quad \pi^k_{ij} = \delta_{ik}
    \delta_{jk}, \quad k = 1, \ldots 4; \nonumber \\
    \Pi_{-} = \frac{1}{4} (\mathbb{I}\otimes \mathbb{I} - \Omega^+), \quad \Pi_{+}
    = \mathbb{I} \otimes \mathbb{I} - \Pi_{-}.
\end{eqnarray}

\subsection{Case 1} It is characterized by $n_+ = 1$, $A = A^T$, $B = A$. We notice that $A
= B$ is equivalent to $n_- = 0$. The commutant is given by
$\mathcal{M} = {\rm span} (\mathbb{I} \otimes \mathbb{I}, \mathbb{I}
\otimes \tilde{\sigma}_i, \tilde{\sigma}_i \otimes
\mathbb{I},\tilde{\sigma}_i \otimes \tilde{\sigma}_i, \Omega^+,
\Delta_i^-)$, and $\mathcal{Z} = \mathcal{M}^{\prime} = {\rm span}
(\mathbb{I} \otimes \mathbb{I}, \tilde{\sigma}_i \otimes
\tilde{\sigma}_i, \mathbb{I} \otimes \tilde{\sigma}_i +
\tilde{\sigma}_i \otimes \mathbb{I})$. The projectors are given by
\begin{eqnarray}\label{eqn01}
    P_1 = \mathcal{R}_i \Pi_1 \mathcal{R}_i, \qquad P_2 = \mathcal{R}_i \Pi_4
    \mathcal{R}_i, \nonumber \\ P_3 = \mathcal{R}_i (\Pi_2 + \Pi_3) \mathcal{R}_i. \quad\qquad
\end{eqnarray}

\subsection{Case 2} It is characterized by $n_+ = 1$, $A \ne A^T$, $B = A$, or rather $n_+ > 1$, $B
= A$. In this case $\mathcal{Z} = \mathcal{M} = {\rm span}
(\mathbb{I} \otimes \mathbb{I}, \Omega^+)$, and there are only two
projectors:
\begin{equation}\label{eqn02}
    P_1 = \Pi_-, \qquad
    P_2 = \Pi_+.
\end{equation}

\subsection{Case 3} It is characterized by $n_+ = n_- = 1$, $A = A^T$, $B = \alpha A$,
$\alpha \in \mathbb{R} \smallsetminus \{ - 1, + 1 \}$. We observe
that $[A, B] = 0$, thus it is possible to choose $\tilde{U} =
\hat{U}$. Moreover, $B = \alpha A$ implies $\tilde{\sigma}_{\xi} =
\hat{\sigma}_{\xi}$ for the index ${\xi}$ such that $\lambda_{\xi}^+
\ne 0$ and $\lambda_{\xi}^- \ne 0$. Finally, $\mathcal{Z} =
\mathcal{M} = {\rm span} (\mathbb{I} \otimes \mathbb{I},
\tilde{\sigma}_i \otimes \tilde{\sigma}_i, \mathbb{I} \otimes
\tilde{\sigma}_i, \tilde{\sigma}_i \otimes \mathbb{I})$, and the
projectors are given by
\begin{equation}\label{eqn03}
    P_j = \mathcal{R}_i \Pi_j \mathcal{R}_i, \qquad j = 1, \ldots,
    4.
\end{equation}

In all the remaining cases $\mathcal{M} = {\rm span} (\mathbb{I}
\otimes \mathbb{I})$, part 1 of Theorem \ref{theo1} applies and the
maximal rank stationary state is unique (if there is one).
Therefore, the aforementioned cases are necessary conditions for the
indirect manipulation of the asymptotic state of the target system
$T$ via the auxiliary system $A$.


\section{Stationary states}
\label{sec3}

We separately explore the non-trivial cases described in Section
\ref{sec2}. Following Theorem \ref{theo1}, if a stationary state
$\rho_0$ whose eigenvalues are all non-vanishing can be found, it is
possible to build the whole family of stationary states
$\rho_s^{\infty}$, by using the projectors $\{P_i; i \}$. Finally,
the corresponding stationary state of the target subsystem can be
obtained from
\begin{equation}\label{eq24}
    \rho_{T}^{\infty} = {\rm Tr}_{A} \, \rho_s^{\infty},
\end{equation}
that is by a partial trace over the degrees of freedom of the
auxiliary system. We consider two different choices for the initial
state $\rho_s (0)$, depending on wether there are initial
correlations or not. As the first choice we take into account the
product state
\begin{equation}\label{eq25}
    \rho_s (0) = \rho_{T} (0) \otimes \rho_{A} (0),
\end{equation}
where $\rho_{T} (0)$ and $\rho_{A} (0)$ are arbitrary states for the
two subsystems, that will be written using a Bloch vector
representation as
\begin{equation}\label{eq26}
    \rho_{T} (0) = \frac{1}{2} \Bigl( \mathbb{I} + \sum_{k = 1}^3 \rho^T_k \sigma_k
    \Bigr),
\end{equation}
with real coefficients $\rho^T_k$, and analogously for $\rho_A (0)$,
with real coefficients $\rho^A_k$. The choice (\ref{eq26}) refers to
initially uncorrelated systems, that will in general couple during
their joint, dissipative evolution, because of the off-diagonal
block $B$ in the Kossakowski matrix $C$. Alternatively, we consider
the pure initial state
\begin{equation}\label{eq27}
     \rho_s (0) = \vert \psi \rangle \langle \psi \vert, \quad \vert
     \psi \rangle = \sqrt{P} \vert \uparrow \rangle \otimes \vert
     \uparrow \rangle + \sqrt{1 - P} \vert \downarrow \rangle
     \otimes \vert \downarrow \rangle,
\end{equation}
where $P \in \mathbb{R}$, and $\vert \uparrow \rangle$, $\vert
\downarrow \rangle$ are the $+1$, respectively $-1$ eigenvalues of
the operator $\sigma_3$. This state is entangled if $P \ne 0, 1$,
and it is maximally entangled if $P = \frac{1}{2}$. It is not an
arbitrary entangled state, nevertheless it can be used to test the
impact of an initial quantum correlation on the manipulation of the
stationary state of $T$.

Although their algebraic structures are different, Cases 1 and 3
lead to the same results. Since $\mathcal{Z} =
\mathcal{M}^{\prime}$, in Case 1 there is not need of $\rho_0$,
whereas the simplest maximal rank stationary state in Case 3 is
given by the maximally mixed state $\rho_0 = \mathbb{I} \otimes
\mathbb{I}$. If there is not correlation in the initial state, it is
not possible to manipulate the stationary state of the system $T$ by
means of the ancilla $A$. In fact, the coefficients of Bloch vector
representation of $\rho_T^{\infty}$, denoted by $\rho^{\infty}_i$,
$i = 1, 2, 3$, depends only on $\rho_T (0)$:
\begin{eqnarray}\label{eq28}
    \rho^{\infty}_1 &=& u_{\xi 1} \Bigl( \rho^T_1 u_{\xi 1} - \rho^T_2 u_{\xi 2}
    + \rho^T_3 u_{\xi 3} \Bigr) \nonumber \\
    \rho^{\infty}_2 &=& - u_{\xi 2} \Bigl( \rho^T_1 u_{\xi 1} - \rho^T_2 u_{\xi 2}
    + \rho^T_3 u_{\xi 3} \Bigr) \\
    \rho^{\infty}_3 &=& u_{\xi 3} \Bigl( \rho^T_1 u_{\xi 1} - \rho^T_2 u_{\xi 2}
    + \rho^T_3 u_{\xi 3} \Bigr), \nonumber
\end{eqnarray}
where $\xi \in \{1,2,3\}$ is such that $\lambda_{\xi}^+ \ne 0$ in
Case 1, $\lambda_{\xi}^+ \ne 0$ and $\lambda_{\xi}^- \ne 0$ in Case
3. If we consider the (possibly entangled) initial state
(\ref{eq27}), the dependence on $P$ is apparent:
\begin{eqnarray}\label{eq29}
    \rho^{\infty}_1 &=& (2P - 1) u_{\xi 1} u_{\xi 3} \nonumber \\
    \rho^{\infty}_2 &=& - (2P - 1) u_{\xi 2} u_{\xi 3} \\
    \rho^{\infty}_3 &=& (2P - 1) u_{\xi 3}^2 \nonumber
\end{eqnarray}
Therefore, at different correlated initial states there correspond
different stationary states $\rho_T^{\infty}$. Manipulations of the
asymptotic states of the target system are possible only when there
is an initial correlation between $T$ and $A$.

In Case 2, since the expression of the stationary state
$\rho_T^{\infty}$ is more involved, we prefer to present a concrete
example in which both uncorrelated and correlated initial states
allow indirect manipulations of the asymptotic states. A simple case
is given by the choice
\begin{equation}\label{eq30}
    A = B = \left[%
\begin{array}{ccc}
  a & i b & 0 \\
  -i b & a & 0 \\
  0 & 0 & a \\
\end{array}%
\right],
\end{equation}
with the condition $a^2 - b^2 \geqslant 0$ expressing the complete
positivity of the evolution. In this case, the maximal rank
stationary state is found to be
\begin{equation}\label{eq30bis}
    \rho_0 = \frac{1}{4} \Bigl( \mathbb{I} \otimes \mathbb{I} + \frac{b}{a}
    (\mathbb{I} \otimes \sigma_3 + \sigma_3 \otimes \mathbb{I}) +
    \Bigl( \frac{b}{a} \Bigr)^2 \sigma_3 \otimes \sigma_3 \Bigr),
\end{equation}
and the asymptotic state of $T$ for the uncorrelated initial state
has components
\begin{eqnarray}\label{eq31}
    \rho^{\infty}_1 &=& 0 \nonumber \\
    \rho^{\infty}_2 &=& 0 \\
    \rho^{\infty}_3 &=& \tau \Bigl( 3 + \sum_{k = 1}^3 \rho^T_k \rho^A_k \Bigr), \nonumber
\end{eqnarray}
where
\begin{equation}\label{eq32}
\tau = \frac{a b}{3 a^2 + 2 b^2}.
\end{equation}
For the correlated initial states we get
\begin{eqnarray}\label{eq33}
    \rho^{\infty}_1 &=& 0 \nonumber \\
    \rho^{\infty}_2 &=& 0 \\
    \rho^{\infty}_3 &=& 4 \tau \Bigl( 1 + \sqrt{P (1 - P)} \Bigr). \nonumber
\end{eqnarray}
Therefore, in both cases it is possible to manipulate
$\rho_T^{\infty}$. We stress that it is possible to drive the
asymptotic state of $T$ through the initial state of $A$ even if the
two system are initially uncorrelated, and there is not a
Hamiltonian coupling between them.


\section{Discussion and Conclusions}
\label{sec4}

We have explored the asymptotic performance of the indirect control
method when both target and auxiliary systems are two-level systems,
and they evolve under a purely dissipative dynamics. We have assumed
that the two systems interact separately with an homogeneous
environment, leading to a particular form of the Kossakowski matrix
$C$ for the composite system. We have found that the conditions
expressed in Case 2 are necessary conditions for the indirect
manipulation of the stationary state of $T$ through the initial
state of $A$ when the initial state is a product state. We have
given a numerical example in which this dependence is explicit. We
have also proved that, in all the non-trivial cases considered, an
initial entanglement between the two systems is effective for
control purposes.

Initial states with a different correlation between the two parties
produce different stationary states for a reduced subsystem whenever
some correlation survives to the decohering action of the
environment. This is also true for more general models than the one
described in this contribution.

For initially uncorrelated states, the dissipative evolution has to
provide the necessary entanglement, that has to be preserved in the
large time limit (for the asymptotic entanglement in a quantum
dynamical semigroup with purely dissipative evolution, see the
results presented in \cite{bena2}). Therefore, the ability of
varying the stationary state of $T$ is a controlled dissipative
mechanism. In this sense, in the indirect control approach the
environmental action can be considered as a resource. This kind of
behavior has already been observed when dealing with accessibility
and controllability of a pair of qubits immersed in a bath of
decoupled harmonic oscillators, in an exactly solvable model
\cite{roma2}. Therefore, it is not an artifact of the Markovian
nature of the evolution.

Eq. (\ref{eq32}) gives a limited ability of manipulation of
$\rho_T^{\infty}$. However, the case here discussed is intended to
represent an example of explicit dependence, not a complete
treatment of the asymptotic reachable set. Moreover, the relevant
case $A = B$ is important in concrete experimental situations, for
example in the study of the resonance fluorescence \cite{arga,puri},
or in the analysis of the weak coupling of two atoms to an external
quantum field \cite{bena3}.


\end{document}